\documentclass[mathleft
]{an}
\usepackage{graphicx}
\usepackage{times}
\def\gtsima{$\; \buildrel > \over \sim \;$}
\def\ltsima{$\; \buildrel < \over \sim \;$}
\def\gtrsim{\lower.5ex\hbox{\gtsima}}
\def\lesssim{\lower.5ex\hbox{\ltsima}}

\begin{document}

\Pagespan{1}{}
\Yearpublication{????}%
\Yearsubmission{2010}%
\Month{????}%
\Volume{????}%
\Issue{????}%

\title{Remnants of massive metal-poor stars: viable engines for ultra-luminous X-ray sources}

\author{M.~Mapelli
\inst{1}\fnmsep\thanks{Corresponding authors:
\email{michela.mapelli@mib.infn.it}\newline}
\and  E.~Ripamonti\inst{1}
\and  L.~Zampieri\inst{2}
\and  M.~Colpi\inst{1}
}
\titlerunning{Remnants of massive metal-poor stars and ULXs}
\authorrunning{Mapelli et al.}
\institute{
Physics Department `G. Occhialini', University of Milano-Bicocca, Piazza della Scienza 3, I--20126 Milan, Italy 
\and 
INAF-Osservatorio astronomico di Padova, Vicolo dell'Osservatorio 5, I--35122, Padova, Italy}

\received{31 August 2010}
\accepted{???}
\publonline{later}


\abstract{%
Massive metal-poor stars might end their life by directly collapsing into massive ($\approx{}25-80\,{}$ M$_\odot{}$) black holes (BHs). We derive the number of massive BHs (N$_{\rm BH}$) that are expected to form per galaxy via this mechanism. We select a sample of 66 galaxies with X-ray coverage, measurements of the star formation rate (SFR) and of the metallicity. We find that N$_{\rm BH}$ correlates with the number of observed ultra-luminous X-ray sources (ULXs) per galaxy (N$_{\rm ULX}$) in this sample. We discuss the dependence of N$_{\rm ULX}$ and of N$_{\rm BH}$ on the SFR and on the metallicity.}

\maketitle

\section{Introduction}
 Most ultra-luminous X-ray sources (ULXs) are located in galaxies with a high star formation rate (SFR, e.g.  Irwin, Bregman \&{} Athey  2004).
 The ULXs match the correlation between X-ray luminosity and SFR reported by various studies (Grimm, Gilfanov \&{} Sunyaev 2003; Ranalli, Comastri \&{} Setti 2003; Gilfanov, Grimm \&{} Sunyaev 2004a,b,c; Kaaret \&{} Alonso-Herrero 2008; Mineo \&{} Gilfanov 2010). Furthermore, the same studies indicate that the luminosity function of ULXs is the direct extension of the function for high-mass X-ray binaries (HMXBs). Recent papers suggest a correlation between ULXs and low-metallicity environments, and propose that this may be connected with the influence of metallicity on the evolution of massive stars (Pakull \&{} Mirioni 2002; Zampieri et al. 2004; Soria et al. 2005; Swartz, Soria \&{} Tennant 2008).
This scenario has been explored in detail by Mapelli, Colpi \&{} Zampieri (2009, hereafter M09), by Zampieri \&{} Roberts (2009) and by Mapelli et al. (2010, hereafter M10), highlighting that a large fraction of ULXs may actually host massive ($\sim{}30-80\,{}{\rm M}_{\odot{}}$) stellar black holes (BHs) formed in a low-metallicity environment.
In fact, low-metallicity ($Z\lesssim{}0.4\,{}Z_\odot{}$) massive stars lose only a small fraction of their mass due to stellar winds (Maeder 1992, hereafter M92; Heger \&{} Woosley 2002, hereafter HW02; Heger et al. 2003, hereafter H03; Belczynski et al. 2010, hereafter B10) and can directly collapse (Fryer 1999; B10) into massive BHs ($25\,{}{\rm M}_\odot{}\le{}m_{\rm BH}\le{}80\,{}{\rm M}_\odot{}$). These massive BHs can power  most of the known ULXs without requiring super-Eddington accretion or anisotropic emission. Furthermore, their formation mechanism can explain the correlation between ULXs and SFR, and the fact that ULXs are preferentially found in low-metallicity  regions.


\section{Sample of galaxies}
In this proceeding, we consider a sample of 66 galaxies. All of them have X-ray coverage, at least one measurement of the star formation rate (SFR) and of the metallicity ($Z$). 64 galaxies are taken from the sample listed in Table~1 of M10. 
 The remaining two are  I~Zw~18 and the interacting pair SBS~0335052/SBS~0335052W\footnote{In this proceeding, we consider the interacting pairs as a unique object, for consistency with M10.}. These two objects are extremely metal-poor galaxies (XMDs, Moiseev, Pustilnik \&{} Kniazev 2010, and references therein) and are important, because they are the only galaxies with $Z<0.03\,{}Z_\odot{}$ and with X-ray observations. Their properties and the corresponding references are listed in Table~1. For details about the data and the properties of the other 64 galaxies, see M10. 

For all the galaxies in the sample, we derive a fiducial value for the SFR (when there is more than one measurement, we take, in general, the average value), for the metallicity (we adopt an uniform calibration, see Pilyugin \&{} Thuan 2005; when a metallicity gradient is available, we take the value of $Z$ at 0.7 Holmberg radii, see M10) and we estimate the number of ULXs N$_{\rm ULX}$ after subtracting the background contamination (see M10 for details).
\begin{table}
\begin{center}
\caption{Properties of the two XMDs in the sample.}
\label{tlab}
\begin{tabular}{cccc}\hline
Galaxy  & SFR [M$_\odot{}$ yr$^{-1}$]$^{\rm a}$ & $Z$ [$Z_\odot{}$]$^{\rm b}$ & N$_{\rm ULX}$$^{\rm c}$ \\ 
\hline
\noalign{\vspace{0.1cm}}
I Zw 18                    & 0.07 & 0.02  & 1\\
SBS 0335-052 & 1.1  & 0.025 & 2.83\\
\noalign{\vspace{0.1cm}}
\hline
\end{tabular}

\footnotesize{$^{\rm a}$ SFR from Wu et al. (2007) for I~Zw~18 and from Thuan, Izotov \&{} Lipovetsky (1997), Pustilnik, Pramskij \& Kniazev (2004) and Johnson, Hunt, Reines (2009) for SBS~0335-052.
$^{\rm b}$ Metallicity from Thuan et al. (2004) for both galaxies.
$^{\rm c}$N$_{\rm ULX}$ is the number of ULXs per galaxy after subtracting the background contamination. Thuan et al. (2004) find 1 ULX in I~Zw~18 and 3 ULXs in SBS~0335-052/SBS~0335-052W. The estimated contamination is 0.00 in I~Zw~18 and 0.17 (upper limit) in SBS~0335-052/SBS~0335-052W.}
\end{center}
\end{table}

\section{Observational results}
\begin{figure}
\begin{center}
\includegraphics[width=50mm]{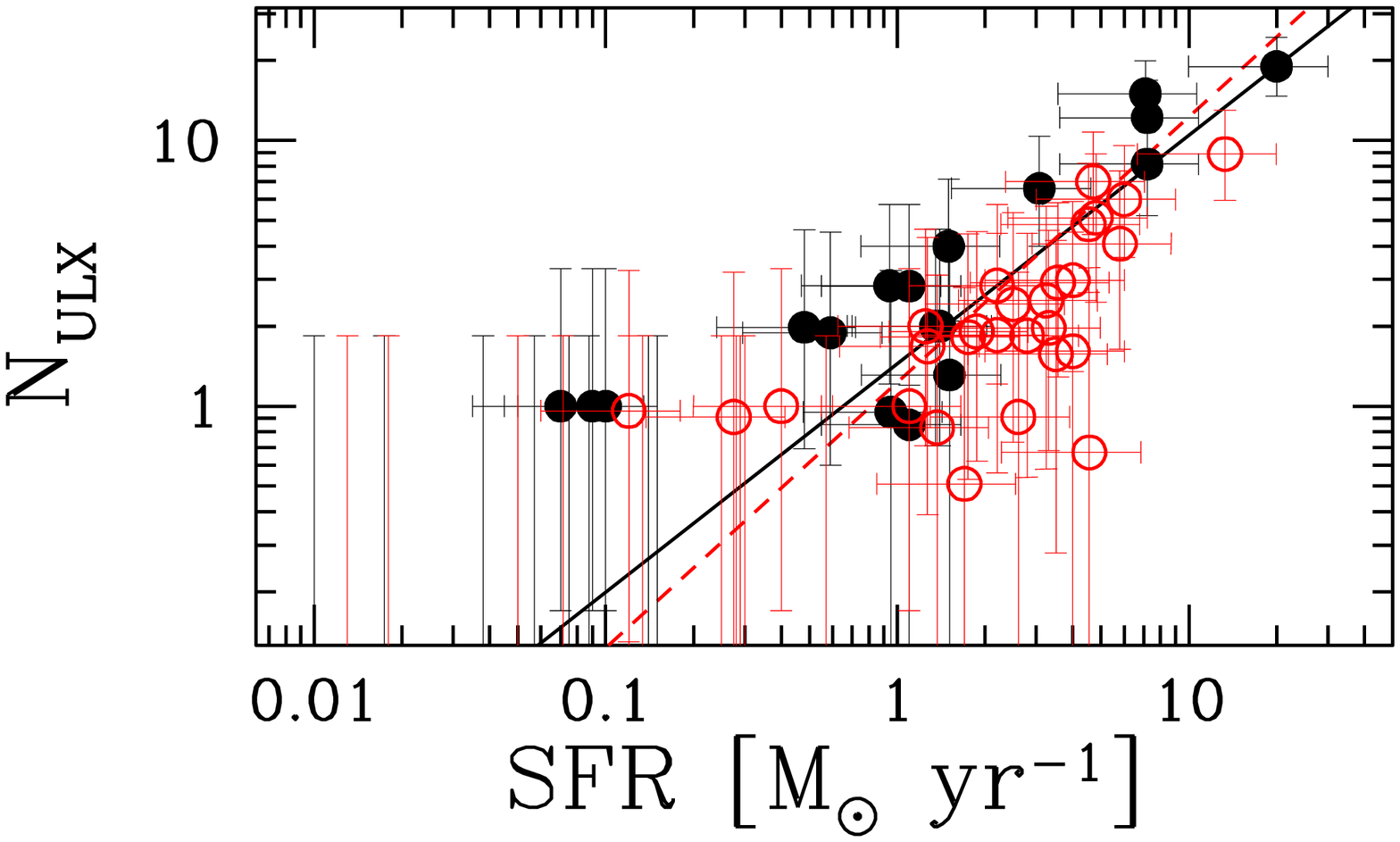}
\includegraphics[width=50mm]{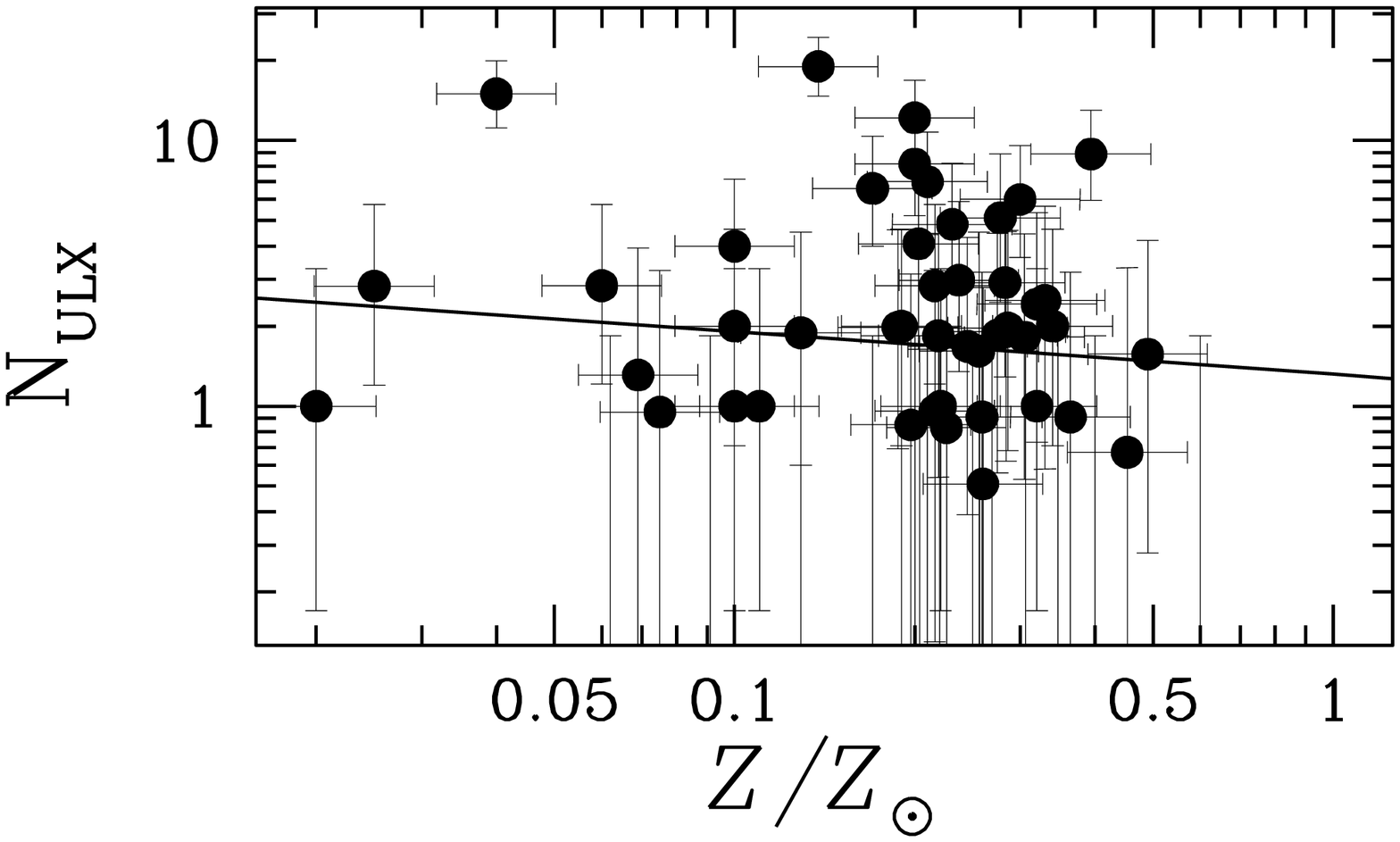}
\includegraphics[width=50mm]{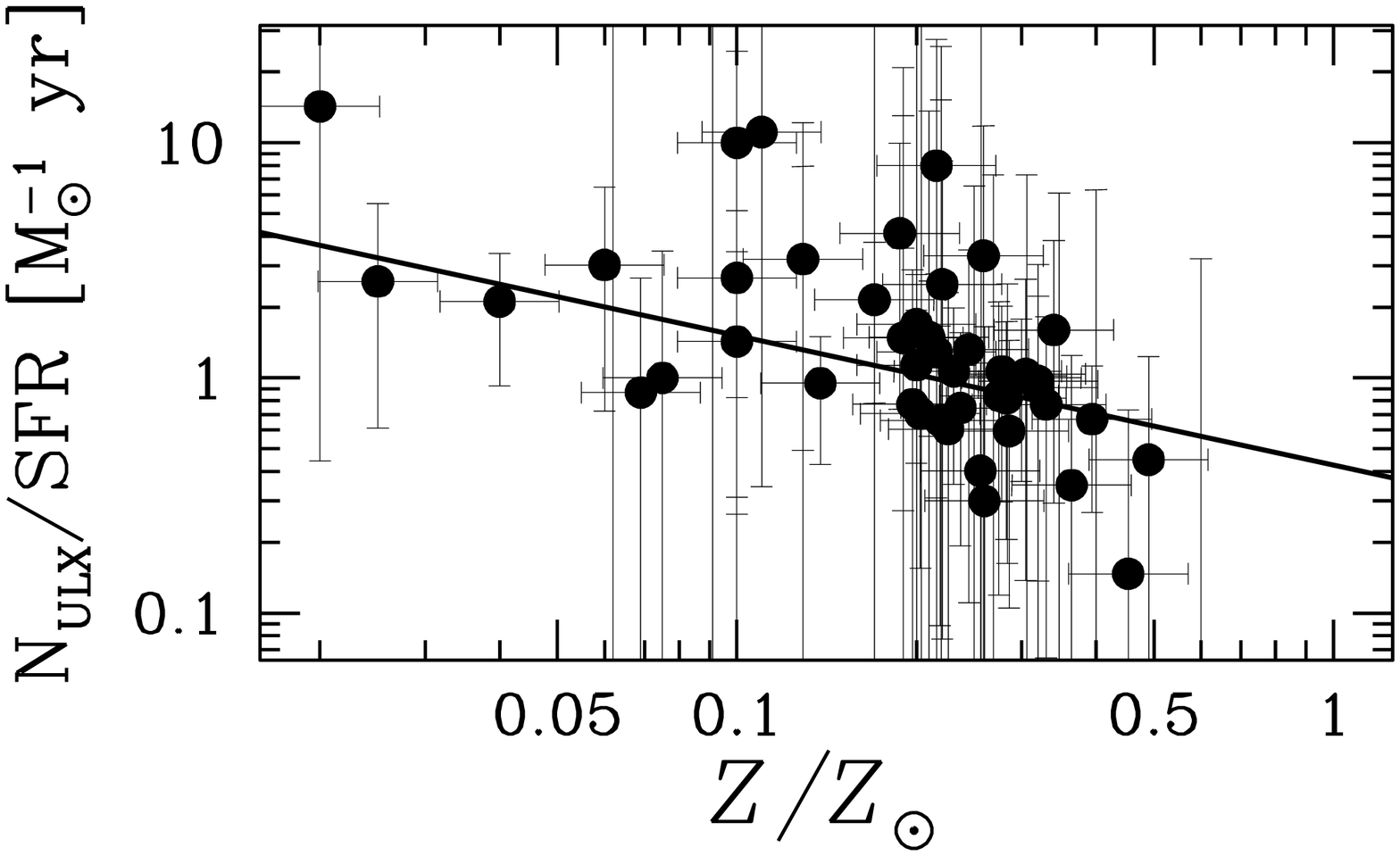}
\caption{
Upper panel: N$_{\rm ULX}$ versus the SFR. Filled black circles: galaxies with metallicity $\le0.2\,{}Z_\odot{}$; open circles (red on the web): galaxies with metallicity $>0.2\,{}Z_\odot{}$. Solid line: power-law fit for the entire sample; dashed line (red on the web): power-law fit obtained assuming that the index of the power law is equal to $1$. 
Central panel: N$_{\rm ULX}$ versus $Z$. Filled black circles: entire sample; solid line: power-law fit. 
Lower panel: N$_{\rm ULX}/{\rm SFR}$ versus $Z$. Filled black circles: entire sample. Solid line: power-law fit. 
In all the panels: the error bars on both the $x-$ and the $y-$ axis are $1\,{}\sigma{}$ errors.}
\label{label1}
\end{center}
\end{figure}
The data collected from the literature were analyzed following the same procedure as described in M10. In particular, we adopt the $\chi{}^2$ analysis (although such method might not  be completely suitable for small samples, see M10), perform the power-law fits and calculate the correlation coefficients (see Table~2).
From the upper panel of Fig.~\ref{label1} it appears that there is a strong correlation between N$_{\rm ULX}$ and the SFR in our sample. Such correlation is consistent with a linear relation (see the best-fitting values reported in Table~2), in agreement with previous studies (see e.g. Grimm, Gilfanov \&{} Sunyaev 2003). Instead, no significant correlation appears between  N$_{\rm ULX}$ and $Z$ (central panel of Fig.~\ref{label1}). However, a marginally significant correlation exists between the number of ULXs normalized to the SFR (N$_{\rm ULX}/$SFR) and the metallicity. This suggests that the metallicity affects the formation of ULXs, but its contribution is less important than that of the SFR. 
The idea that metallicity plays a role in the origin of ULXs is consistent with previous observations (see e.g. Swartz et al. 2008 and references therein) and with some recent theoretical models (M09; Zampieri \&{} Roberts 2009; Linden et al. 2010; M10).
Finally, the presence of the two XMDs in our sample does not significantly change the best-fitting values (Table~2) with respect to those derived in M10.

\section{Comparison of the data with the theoretical model}
In this Section, we analyze the observational data collected from the literature on the light of the theoretical model recently proposed by M09 and M10. First, we briefly summarize such model.

\subsection{Theoretical model}
According to numerical calculations (Fryer 1999; HW02; H03), a star that, at the end of its life, has a final mass $m_{\rm fin}\ge{}40\,{}{\rm M}_\odot{}$ is expected to directly collapse into a BH. In this case, the mass of the remnant BH is likely  more than half of the final mass of the progenitor star, as  relatively small mass ejection is expected in the direct collapse. Thus, stars that at the end of their lives have $m_{\rm fin}\ge{}40\,{}{\rm M}_\odot{}$ are likely to produce massive BHs (B10).
The final masses of the stars strongly depend on their metallicity. Massive stars with metallicity close to solar cannot have final masses larger than $m_{\rm fin}\sim{}10-15\,{}{\rm M}_\odot{}$, even if their initial mass was very large, as they are expected to lose a lot of mass due to stellar winds (H03). Instead, massive stars with lower metallicity are less affected by stellar winds, and retain a larger fraction of their initial mass. If its metallicity is sufficiently low, a star can have a final mass $m_{\rm fin}\ge{}40\,{}{\rm M}_\odot{}$ and can directly collapse into a massive BH with a mass $25\,{}{\rm M}_\odot{}\le{}m_{\rm BH}\le{}80\,{}{\rm M}_\odot{}$ (HW02; H03; B10).
The recent model by B10 accounts for the fact that stars with $m_{\rm fin}\ge{}40\,{}{\rm M}_\odot{}$ can directly collapse into BHs. For this reason, in this model BHs with mass as large as $80\,{}{\rm M}_\odot{}$ are allowed to form.


On the basis of this scenario, we can derive the expected number of massive BHs per galaxy (${\rm N}_{\rm BH}$) as a function of the star formation rate (SFR) and of the metallicity $Z$ (see M09, M10):
\begin{equation}\label{eq:totnum}
{\rm N}_{\rm BH}({\rm SFR},\,{}Z)=A({\rm SFR})\,{}\,{}\,{}\int_{m_{\rm prog}(Z)}^{m_{\rm max}}m^{-\alpha{}}\,{}{\rm d}m,
\end{equation}
where $m_{\rm max}$ is the maximum stellar mass (we assume $m_{\rm max}=120\,{}{\rm M}_\odot{}$) and  $\alpha{}$ is the  slope  of the initial mass function (IMF). Here, we adopt the Kroupa IMF, for which $\alpha{}=1.3$ if $m\le0.5\,{}{\rm M}_\odot{}$ and $\alpha{}=2.3$ for larger masses (Kroupa 2001).
$m_{\rm prog}(Z)$ is the minimum initial stellar mass (i.e. the mass at zero-age main sequence) for which a star is the progenitor of a massive BH. As we discussed above, $m_{\rm prog}(Z)$ strongly depends on the metallicity. In our calculations, we assume $m_{\rm prog}(Z)$ to be the initial stellar mass for which the mass of the remnant is $m_{\rm BH}\gtrsim{}25\,{}{\rm M}_\odot{}$, according to the model by B10\footnote{M10 also consider an alternative model by Portinari, Chiosi \&{} Bressan 1998.}.

 Finally, $A({\rm SFR})$, the normalization constant in equation~(\ref{eq:totnum}), can be estimated as
\begin{equation}\label{eq:norm1}
A({\rm SFR})=\frac{{\rm SFR}\,{}\,{}\,{}t_{\rm co}}{\int_{m_{\rm min}}^{m_{\rm max}}m^{1-\alpha{}}\,{}{\rm d}m}, 
\end{equation}
where $m_{\rm min}$ is the minimum stellar mass (we assume $m_{\rm min}=0.08$ ${\rm M}_\odot{}$), SFR is the current star formation rate and $t_{\rm co}$ is the characteristic lifetime of a possible companion of the massive BH. In fact, we are not interested in all the massive BHs, but only in those that could acquire massive stellar companions and power observable ULXs. In this proceeding, we adopt a constant value $t_{\rm co}=10^7$ yr, which is the lifetime of a $\sim{}15\,{}{\rm M}_\odot$ star. 
\begin{figure}
\begin{center}
\includegraphics[width=50mm]{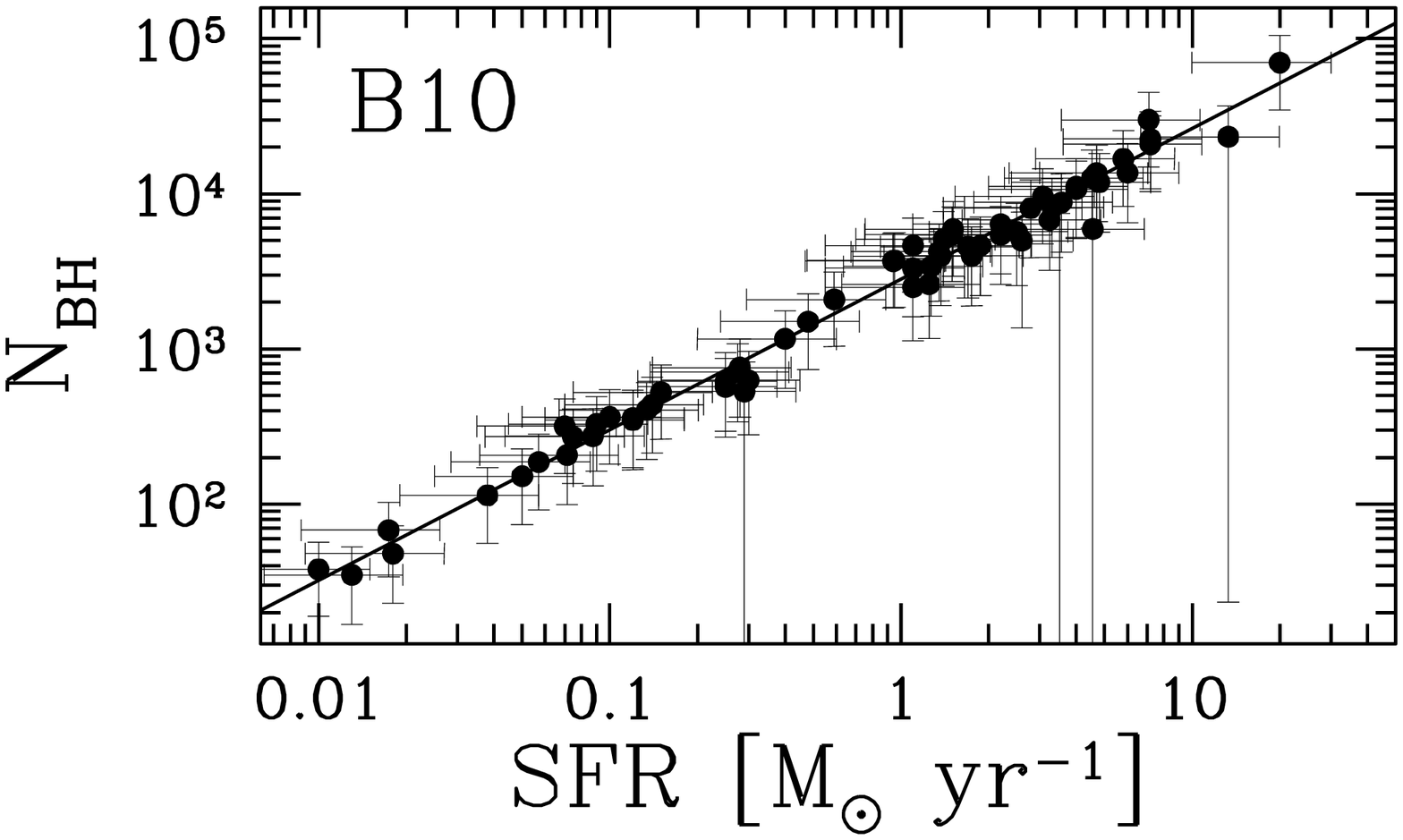}
\includegraphics[width=50mm]{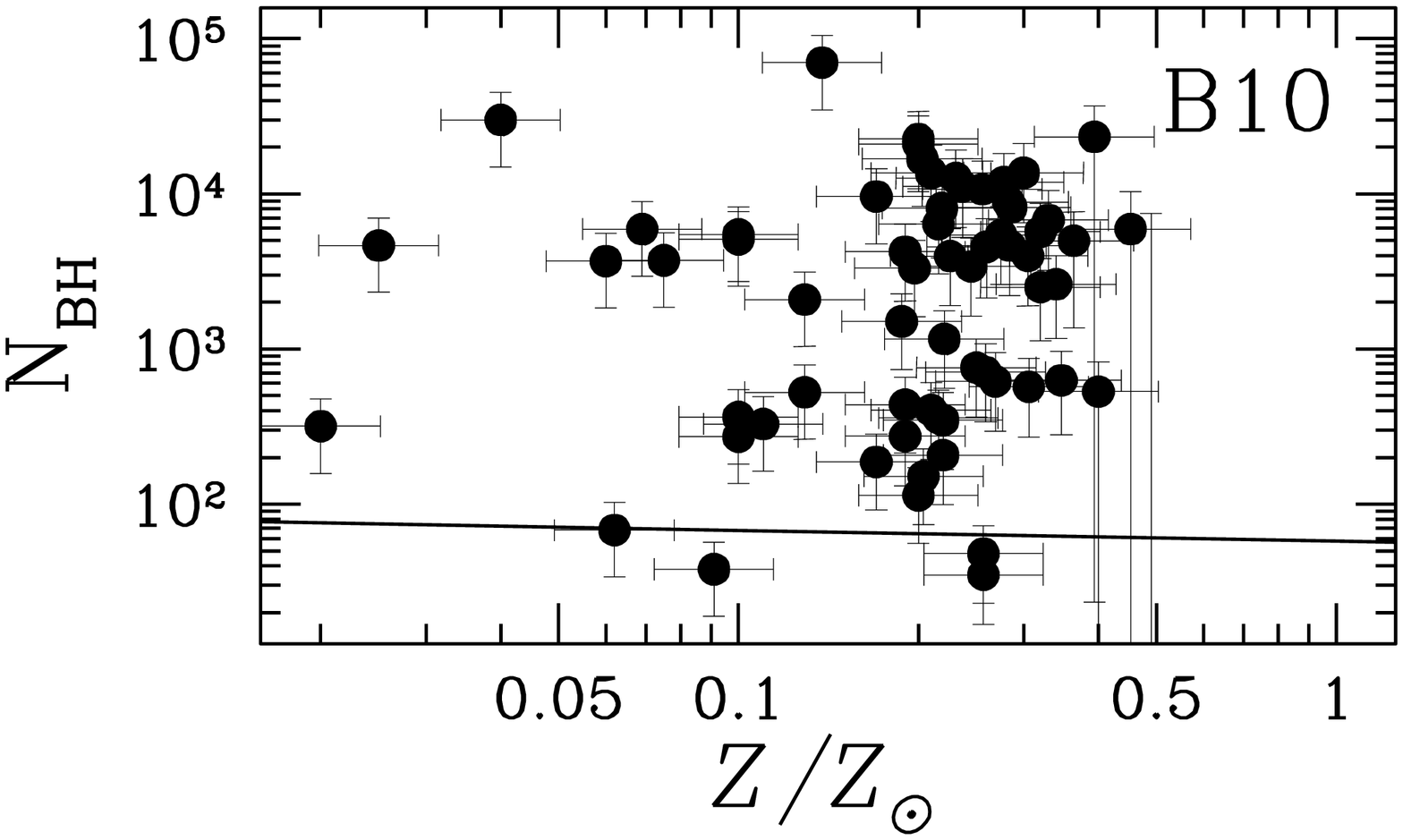}
\includegraphics[width=50mm]{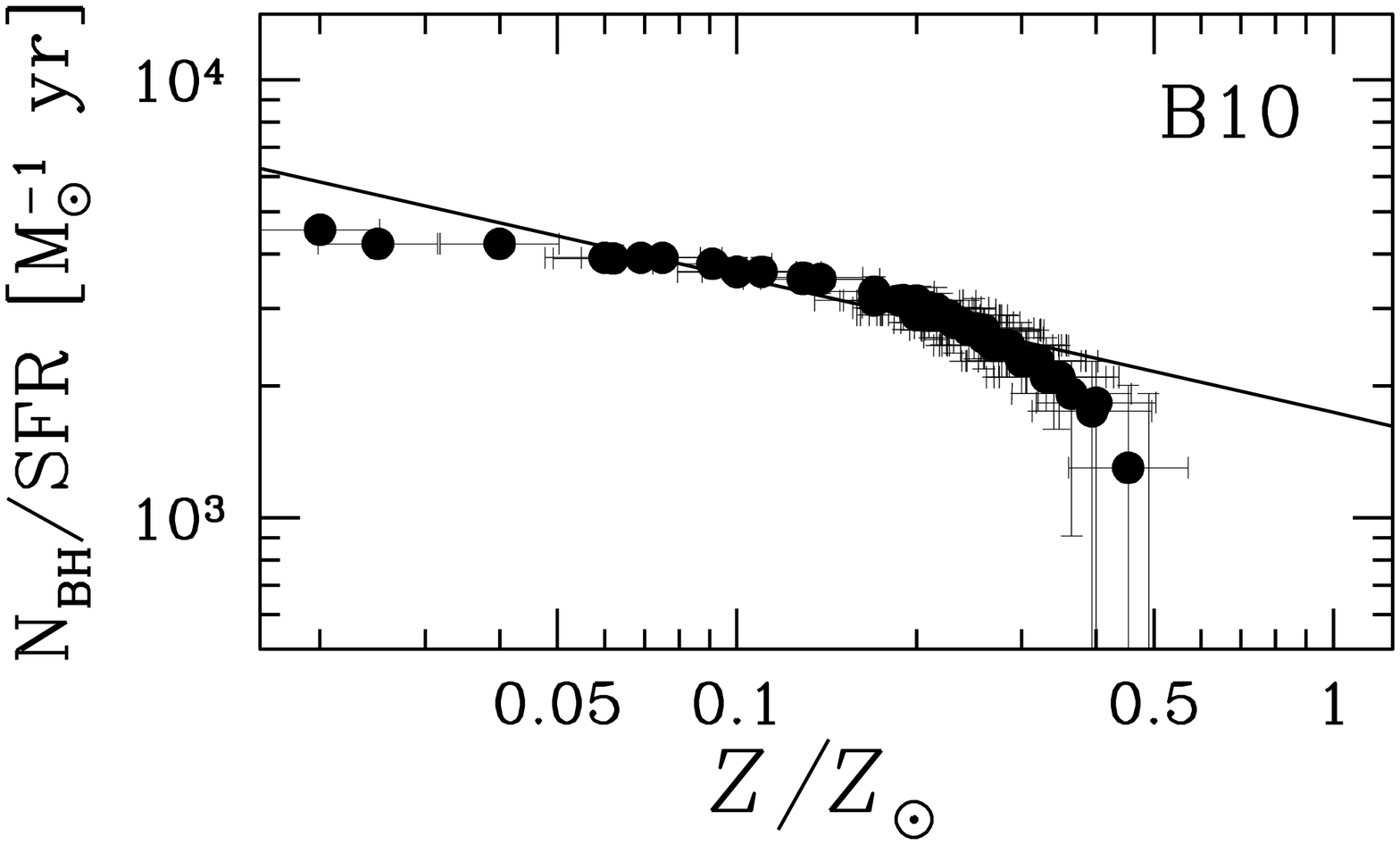}
\caption{
Upper panel: N$_{\rm BH}$ (derived using the model from B10) versus the SFR. 
Central panel: N$_{\rm BH}$ versus $Z$. 
Lower panel: N$_{\rm BH}/{\rm SFR}$ versus $Z$. 
In all the panels, filled black circles: entire sample; solid line: power-law fit for the entire sample; the error bars on both the $x-$ and the $y-$ axis are $1\,{}\sigma{}$ errors.
}
\label{label2}
\end{center}
\end{figure}
\begin{figure}
\begin{center}
\includegraphics[width=50mm]{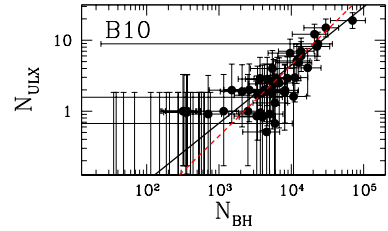}
\caption{
 Number of observed ULXs per galaxy N$_{\rm ULX}$ versus the number of expected massive BHs per galaxy N$_{\rm BH}$, derived using the model from B10. The solid line is the power-law fit for the entire sample. The dashed line (red on the web) is the power-law fit obtained assuming that the index of the power law is $=1$.
The error bars on both the $x-$ and the $y-$ axis are $1\,{}\sigma{}$ errors. }
\label{label3}
\end{center}
\end{figure}

\subsection{Results}
Fig.~\ref{label2} shows the behaviour of the theoretical model when applied to the observed SFR and metallicity. As assumed in the model, N$_{\rm BH}$ scales linearly with the SFR. The central panel of Fig.~\ref{label2} shows that there is no significant correlation between N$_{\rm BH}$ and $Z$, although we imposed, in the model, that N$_{\rm BH}$ does depend on $Z$. This absence of correlation agrees with what we found for N$_{\rm ULX}$ versus $Z$ (central panel of Fig.~\ref{label1}).  Finally, only when the effect of the SFR is subtracted (by normalizing N$_{\rm BH}$ to the SFR, lower panel of Fig.~\ref{label3}), it is possible to see the dependence of N$_{\rm BH}$ on the metallicity. The behaviour of N$_{\rm BH}/$SFR versus $Z$ in the model is consistent with that of N$_{\rm ULX}/$SFR versus $Z$ in the data.

Fig.~\ref{label3} and Table~2 indicate that there is a correlation between  N$_{\rm BH}$ and  N$_{\rm ULX}$. This correlation is slightly more significant than that between N$_{\rm ULX}$ and SFR, when considering both the $\chi^2$ analysis and the correlation coefficient. 

Recently, Linden et al. (2010) proposed a different model to explain the connection between low-metallicity environments and ULXs. They indicate that the number, the lifetime and (less significantly) the luminosity of HMXBs are enhanced by low metallicity.
 Linden et al. (2010) also point out a possible problem of M10's model: massive BHs born via direct collapse likely do not receive a natal kick and this fact excludes, in the model by Linden et al. (2010),  the possibility of forming a HMXB via Roche lobe overflow (RLO).

On the other hand, Linden et al. (2010) always require super-Eddington emission, to explain the ULXs. Furthermore, the process of  the direct collapse and the physics of the binaries which include massive BHs born from it are still far from being understood. For example, natal kicks might still be present, due to asymmetries induced by sterile neutrinos (e.g. Kusenko 2006). In alternative, kicks might occur for different reasons, e.g. due to three-body encounters in the parent cluster (Mapelli et al., in preparation). Such scenario might also explain why ULXs are often found to be displaced with respect to star-forming regions (e.g. Swartz, Tennant \&{} Soria 2009).

\section{Conclusions}
In this proceeding, we considered a sample of 66 galaxies. All of them have X-ray coverage, at least one measurement of the SFR and of $Z$. This sample includes two XMDs, which have extremely low metallicity and host a relatively high number of ULXs, when compared to their SFR.

We find that there is a strong correlation between the number of ULXs per galaxy (N$_{\rm ULX}$) and the SFR. This is consistent with previous studies (e.g. Grimm, Gilfanov \&{} Sunyaev 2003). We also find a marginally significant anti-correlation between N$_{\rm ULX}/$SFR and the metallicity.
This might indicate that the metallicity is the missing ingredient, to understand the formation of ULXs, although the error bars are still very large and the sample of galaxies is quite small.

Recently, M09 and M10 suggested that ULXs might be connected with massive BHs ($25-80\,{} M_\odot{}$) formed by the direct collapse of massive metal-poor stars (Fryer 1999; B10). We derive the number of BHs per galaxy (N$_{\rm BH}$) predicted by M10 and compare it with the observed N$_{\rm ULX}$ in our sample. We find a strong correlation between N$_{\rm BH}$ and N$_{\rm ULX}$.

We note that the model by B10 derives the mass of the remnant for single stars only, without considering stars in binaries. Stars in close binaries likely have a different mass-loss history. Accounting for binary progenitors might strengthen the dependence of the BH mass on metallicity. Therefore, it will be necessary to account for binary evolution, to refine the model of massive BH formation.

Furthermore, 
 the physics of the  direct collapse and the properties of massive BHs born from it are only poorly known. These need to be investigated, in order to understand the process of mass transfer (and of X-ray emission) in binaries including massive BHs.

Finally, we need more observational data, especially measurements of the metallicity, to strengthen our conclusions. XMDs are particularly interesting, because of their peculiarly low metallicity. 

\begin{table}
\begin{center}
\caption{Parameters of the  power-law fits and $\chi{}^2$.}
\label{tlab2}
\begin{tabular}[!h]{llllll}
\hline
$x$
& $y$
& {\scriptsize Index} $^{\rm a}$
& {\scriptsize Normalization}
& $\chi{}^2$ $^{\rm b}$
& r $^{\rm c}$\\
\hline  
\noalign{\vspace{0.1cm}}
{\scriptsize ${\rm N}_{\rm BH}$}  & {\scriptsize ${\rm N}_{\rm ULX}$} & $0.82^{+0.18}_{-0.12}$ &   -2.63$^{+0.48}_{-0.72}$ & $12.0$   & 0.92 \\   
{\scriptsize ${\rm N}_{\rm BH}$}  & {\scriptsize ${\rm N}_{\rm ULX}$} & $1.00$          &   -3.35$^{+0.06}_{-0.06}$ & $13.0$   & 0.92 \vspace{0.2cm}\\   
{\scriptsize SFR}   & {\scriptsize ${\rm N}_{\rm ULX}$}         &  $0.86^{+0.22}_{-0.14}$ &  $0.16^{+0.09}_{-0.12}$  & $19.3$   & 0.88 \\
{\scriptsize SFR}   & {\scriptsize ${\rm N}_{\rm ULX}$}         &  $1.00$                 &  $0.09^{+0.06}_{-0.06}$         &   $19.8$ & 0.88 \vspace{0.2cm}\\

{\scriptsize $Z$}  & {\scriptsize ${\rm N}_{\rm ULX}$} & -0.16$^{+0.28}_{-0.28}$ &  $0.12^{+0.20}_{-0.20}$  &    $86.5$ & -0.15 \\

{\scriptsize $Z$}  & {\scriptsize ${\rm N}_{\rm ULX}/{\rm SFR}$} & -0.55$^{+0.21}_{-0.19}$ &  -0.37$^{+0.16}_{-0.16}$  &    $11.0$  & -0.38\\


\noalign{\vspace{0.1cm}}
\hline
\end{tabular}

\footnotesize{
The SFR and the $Z$ used by the fitting procedure are in units of M$_\odot{}$ yr$^{-1}$ and of $Z_\odot{}$, respectively.
$^{\rm a}$  When the index is equal to 1.00 or to 0.00, without error, it means that it has been fixed to such value.
$^{\rm b}$ $\chi{}^2$ is the non-reduced $\chi{}^2$. The number of degrees of freedom (dof) is 65  when the index has been fixed, 64 in the other cases.
$^{\rm c}$  r is the Pearson correlation coefficient. 
}
\end{center}
\end{table}

\acknowledgements We thank M.~Gomitoni, V.~Andreoni, A.~Bressan, P.~Marigo, the organizers and the participants to the conference ``Ultra-Luminous X-ray sources and Middle Weight Black Holes'' (Madrid, 24th-26th May 2010) for useful discussions. LZ and MC acknowledge financial support through INAF grant PRIN-2007-26.





\begin{thebibliography}{}
\bibitem{}Belczynski, K., Bulik, T., Fryer, C. L., Ruiter, A., Valsecchi, F., Vink, J. S., Hurley, J. R.: 2010, ApJ, 714, 1217 (B10)
\bibitem{}Fryer, C. L.: 1999, ApJ, 522, 413
\bibitem{}Gilfanov, M., Grimm, H.-J., Sunyaev, R.: 2004a, MNRAS, 347L, 57
\bibitem{}Gilfanov, M., Grimm, H.-J., Sunyaev, R.: 2004b, Nuclear Physics B Proceedings Supplements, 132, 369  
\bibitem{}Gilfanov, M., Grimm, H.-J., Sunyaev, R.: 2004c, MNRAS, 351, 1365
\bibitem{}Grimm, H.-J., Gilfanov, M., Sunyaev, R.: 2003, MNRAS, 339, 793
\bibitem{}Heger, A., Fryer, C.L., Woosley, S.E., Langer, N., Hartmann, D.H.: 2003, ApJ, 591, 288 (H03)
\bibitem{}Heger, A., Woosley, S.E.: 2002, ApJ, 567, 532 (HW02)
\bibitem{}Irwin, J. A., Bregman, J. N., Athey, A. E.: 2004, ApJ, 601L, 143
\bibitem{}Johnson, K. E., Hunt, L. K., Reines, A. E.: 2009, AJ, 137, 3788
\bibitem{}Kaaret, P., Alonso-Herrero, A.: 2008, ApJ, 682, 1020
\bibitem{}Kroupa P.: 2001, MNRAS, 322, 231
\bibitem{}Kusenko, A.: 2006, Physical Review Letters, 97, 24, 1301
\bibitem{}Linden, T., Kalogera, V., Sepinsky, J. F., Prestwich, A., Zezas, A., Gallagher, J.: 2010, ApJ, submitted, arXiv:1005.1639
\bibitem{}Maeder, A.: 1992, A\&{}A, 264, 105 (M92)
\bibitem{}Mapelli, M., Colpi, M., Zampieri, L.: 2009, MNRAS, 395L, 71 (M09)
\bibitem{}Mapelli, M., Ripamonti, E., Zampieri, L., Colpi, M., Bressan, A.: 2010, MNRAS, accepted, arXiv:1005.3548 (M10)
\bibitem{}Mineo, S., Gilfanov, M.: 2010, these proceedings
\bibitem{}Moiseev, A. V., Pustilnik, S. A., Kniazev, A. Y.: 2010, MNRAS, 405, 2453
\bibitem{}Pakull, M. W., Mirioni, L.: 2002, astro-ph/0202488
\bibitem{}Pilyugin, L. S., Thuan, T. X.: 2005, ApJ, 631, 231 [PT05]
\bibitem{}Portinari, L., Chiosi, C., Bressan, A.: 1998, A\&{}A, 334, 505
\bibitem{}Pustilnik, S. A., Pramskij, A. G., Kniazev, A. Y.: 2004, A\&{}A, 425, 51
\bibitem{}Ranalli, P., Comastri, A., Setti, G.: 2003, A\&{}A, 399, 39
\bibitem{}Soria, R., Cropper, M., Pakull, M., Mushotzky, R., Wu, K.: 2005, MNRAS, 356, 12
\bibitem{}Swartz, D. A., Soria, R., Tennant, A. F.: 2008, ApJ, 684, 282
\bibitem{}Swartz D. A., Tennant A. F., Soria R., 2009, ApJ, 703, 159
\bibitem{}Thuan, T. X., Izotov, Y. I., Lipovetsky, V. A.: 1997, ApJ, 477, 661 
\bibitem{}Thuan, T. X., Bauer, F. E., Papaderos, P., Izotov, Y. I.: 2004, ApJ, 606, 213
\bibitem{}Wu, Y., Charmandaris, V., Hunt, L. K., Bernard-Salas, J., Brandl, B. R., Marshall, J. A., Lebouteiller, V., Hao, L., Houck, J. R.: 2007, ApJ, 662, 952
\bibitem{}Zampieri, L., Mucciarelli, P., Falomo, R., Kaaret, P., Di Stefano, R., Turolla, R., Chieregato, M., Treves, A.: 2004, ApJ, 603, 523
\bibitem{}Zampieri, L., Roberts, T.: 2009, MNRAS, 400, 677
\end{thebibliography}
\end{document}